# All-Indoor Optical Customer Premises Equipment for Fixed Wireless Access

Dominic Schulz[1], Julian Hohmann[1], Peter Hellwig[1], Christoph Kottke[1], Ronald Freund[1], Volker Jungnickel[1], Ralf-Peter Braun[2], Frank Geilhardt[2]

[1]*Fraunhofer Heinrich Hertz Institute (HHI), 10587 Berlin, Germany*
[2]*Deutsche Telekom AG, 10781 Berlin, Germany*
dominic.schulz@hhi.fraunhofer.de

**Abstract:** We demonstrate an LED-based optical wireless link for fixed wireless access applications, at data rates of 1.5 Gbit/s over 50 m. Transmission between indoor equipment and outdoor access point is possible through metal-coated insulation glass. - *Preprint* -

**1. Introduction and Innovation**
The ever-increasing requirements for bandwidth intensive applications, in addition to new features for the Internet of Things (IoT), demand for a next generation of fixed and mobile broadband networks. New applications like smart factories or the progressing digitalization of the society increase the specifications for data rate per area by a factor of 1000 and the number of connected devices by a factor of 100 in the future [1]. A densification of radio cells in mobile networks in combination with the deployment of fiber-to-the-home (FTTH) are considered to enable those next generation networks. FTTH deployments make high data rates and robustness possible but the installations, especially for digging the fiber to connect each individual home on the last hundred meters, are customer specific, costly and time-consuming as well. Especially in urban areas, installation of fiber can be difficult in terms of cost and time. Legal frameworks, e.g. the way of right, or constructional barriers, like railway tracks or rivers, do not allow for fiber installations in some areas. An alternative is the fixed wireless access (FWA), where the last hop to the customer is realized with a wireless link. Instead of connecting each household with a fiber, the fiber will only be deployed to the curb, e.g. to the street lights in a residential area, and the final hop to the building is less costly using so called wireless-to-the-home (WTTH) concepts. With a low-cost wireless technology, WTTH enables an economic broadband expansion, serving the current and upcoming demands for networks in the future [2].

Directed radio links in the millimeter wave region (e.g. 60 GHz) could be used for FWA applications. But due to the already overcrowded spectrum, rare radio bandwidth is also needed for mobile access. To avoid complex coordination between FWA and mobile access applications, we propose to use LED-based optical wireless communication (OWC) to relieve the millimeter wave spectrum and offload traffic to the optical spectrum.

LED-based OWC, also denoted as LiFi, is frequently discussed in recent years. While features like mobility and multi-user support are provided for mobile access, LiFi can also be used as an alternative for 60 GHz in WTTH applications. The probability of a free line-of-sight (LOS) is rather high in urban areas [3]. The required protocols for LiFi in the point-to-point (P2P) topology are simplified. OWC frontends are equipped with additional lenses to obtain a highly directed LOS link and to reduce the geometrical loss. LiFi has been demonstrated for long-range P2P outdoor communication over distances up to 200 m [4,5]. By means of closed-loop rate adaptation, OWC links achieve availabilities of more than 99.9%, despite the fluctuation of the signal-to-noise ratio (SNR) due to the attenuation and scattering in the atmosphere [5].

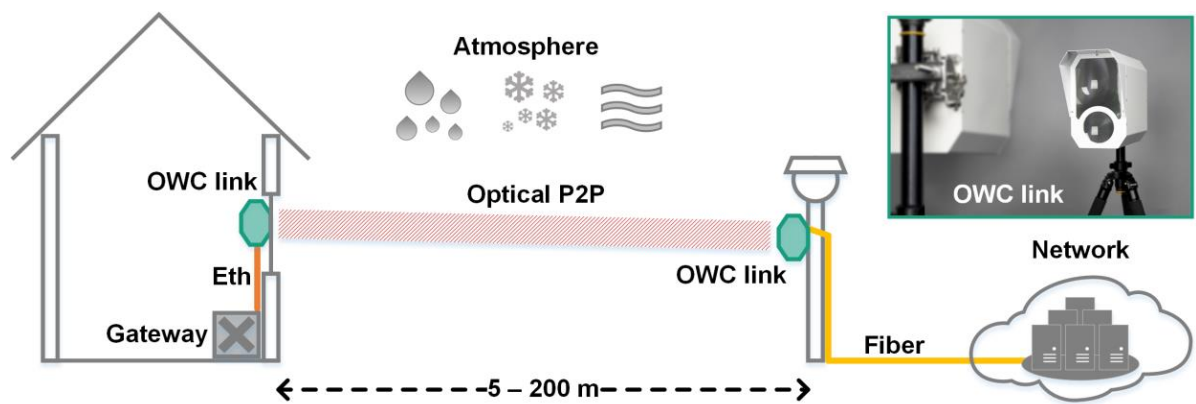

Fig. 1 - Wireless-to-the-Home scenario as an example for Fixed Wireless Access applications; top right: OWC outdoor P2P link

The Innovation Action "Enhanced Lighting for the Internet of Things - (ELIoT)" aims at enhancing the maturity of LiFi, add several new features and enabling LiFi to serve the increased number of users and satisfy the demand of the future IoT. ELIoT will demonstrate LiFi in a wide range of future IoT use cases, one of which is e.g. FWA. With the focus on high data rates, low latencies, secure encrypted transmissions, and increased resilience, ELIoT aims to develop the next generation of wireless networks by adding the use of light as a complement to radio. In the ELIoT project, a previously reported OWC outdoor P2P link has been further developed. Now it offers higher data rate over longer transmission distances, has a more economic housing to be used as an low-cost FWA solution. One particular requirement was to deploy the customer premises equipment (CPE) all indoors. Radio links are entirely blocked by heat insulation glass what implies that the system needs to be split into an indoor and an outdoor unit, connected by a cable to be deployed by the house owner, which is not desired. Telecoms are interested in all-indoor CPE. In this paper, we demonstrate an advanced OWC prototype for FWA applications and report initial measurements in the intended deployment scenario. We report the data rate over several transmission distances and compare the effect of metal-coated heat insulation glasses, which are increasingly used in low-carbon footprint houses, on the transmission performance with a commercial 60 GHz radio link [6].

## 2. Link Design

The developed OWC link enables real-time transmission by using a commercially available digital signal processing (DSP) chipset based on the ITU-T recommendation G.9991 using a bandwidth of 200 MHz. With a direct-current biased orthogonal frequency division multiplexing (DC-OFDM) waveform up to 12 bit/s/Hz spectral efficiency a peak data rate of 2 Gbit/s can be achieved in time-division duplex mode. An Ethernet interface is used for the supply of data. Medium access and phyiscal layer functions as well as digital-to-analog and analog-to-digital conversion are performed by the chipset. Analogue signals are fed into an optical frontend (OFE), comprising transmitter and receiver circuitry. For optical transmission, an LED SFH 4780S from OSRAM with a wavelength of 820 nm and a die size of 0.75x0.75 mm² is used. A half angle of 10° is achieved by an optical concentrator being directly attached to the LED. To increase the LED modulation bandwidth and exploit the full bandwidth of the DSP chip, a sophisticated LED driver is used. A large area Si-PIN photodiode (PD) S6968 from Hamamatsu is combined with a sensitive photoreceiver. To overcome longer distances, the divergence angle of the OWC link and thereby the optical loss is reduced. Due to the large areas of the LED and PD, the focusing requirements for the optics is reduced so that low-cost Fresnel lenses can be used at transmitter and receiver side. The transmitter lens has a focal length of 208 mm and an area of 50 cm² while the receiver lens has the same focal length but an area of 150 cm². With implementing the aforementioned optics, the overall transmitter divergence half angle is reduced to 0.41° and the effective receiver area is increased significantly. At 100 m distance, a spot size of 1.4 m is reached, which makes the link robust against vibrations without the need for costly active tracking. Since closed-loop rate adaptation is implemented, no link margin is used, i.e. impairments due to the weather conditions will reduce the data rate. However, the probability of low atmospheric losses is high [5], due to the short distances in the FWA scenario, i.e. high data rates can be achieved and the bandwidth is used efficiently.

## 3. Measurement Results and Demo

The printed circuit board, comprising the DSP chipset and the OFE, was set up in a waterproof housing, shown in Fig. 1, top right, to withstand outdoor conditions like precipitation. The modules can be mounted at a lamp post with a tip-and-tilt stage that is directly mounted to the housing, for easy alignment and best connectivity. To validate the link performance in the envisioned FWA use case, initial outdoor measurements were performed. At first, the gross data rates were measured at several distances, which are relevant for the FWA application. Therefore, two OWC terminals were aligned to each other with a free LOS. For each distance, the alignment is adapted for the best possible SNR to maximize the data rate. The data rate versus transmission distance is shown in Fig. 2 a). Results for a previous OWC link [5] with 100 MHz bandwidth are given for comparison. Achieved data rates are 1500, 1400 and 1100 Mbit/s at 25, 50 and 100 m transmission distance, respectively, representing an increase of a factor of 2, 2.5 and 2.2 compared to the previous design. The significantly higher data rate is attributed to the enhanced signal bandwidth from 100 to 200 MHz and the new OFE with higher optical output power and larger LED area.

In Fig. 2 b) the effect of inserting a metal-coated double heat insulation glass into the LOS of the link is shown to emulate an indoor CPE installation behind a window. The SNR for the OWC link and the channel-to-interference-and-noise ratio (CINR) for a commercial P2P 60 GHz link from Siklu is shown, respectively. For the OWC link, the SNR is given per carrier while the 60 GHz link provides only an average value. Therefore, the CINR is a constant line. It can be observed that the average SNR drop of the optical link is 13 dB in the electrical domain when comparing free LOS transmission with the inserted insulation glass. This drop is attributed to the fact that light is partially reflected when passing through the heat insulation glass.

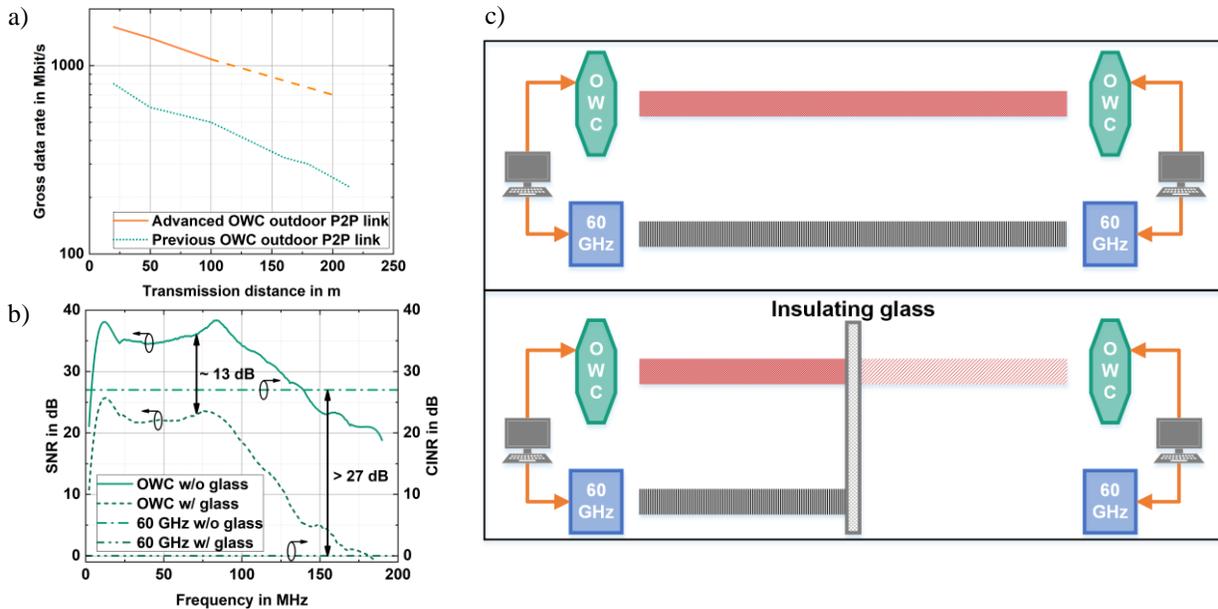

Fig. 2 – a) Measured gross data rate over transmission distance for a previous version and the current advanced version of the OWC link. b) Measured SNR for the current OWC link and CINR for a commercial 60 GHz link with and without the impairment of a coated double-insulation glass. c) Demo concept with the initial free LOS between OWC and 60 GHz link as well as the blocked LOS by a coated insulation glass.

The same experiment was also done with the 60 GHz link. With a free LOS between the two 60 GHz terminals the CINR is 27 dB but when the insulation glass is inserted into the LOS, the CINR drops to 0 dB and communication is interrupted. Heat insulation glass is coated with a thin metal layer, depleting the link margin of the 60 GHz link entirely. Even though the SNR of the OWC link was reduced, too, communication was possible at 600 Mbit/s.

The concept of the demo, shown in Fig. 2 c), is to demonstrate the aforementioned experiment in real-time, highlight the potential of low-cost optical P2P links for FWA applications and demonstrate the new observation that the CPE can be fully deployed indoors behind the heat insulation glass. This makes it easier for telecom operators to deploy FWA links, because no outdoor antenna is needed would require a cable link to the indoor CPE.

The initial situation is a free LOS between OWC and 60 GHz link, with a PC connected to both technologies, monitoring the SNR (OWC) and CINR (60 GHz) as well as the data rate. A coated double-insulation glass will be inserted into the LOS, representing an impairment for both FWA links. It will reduce the SNR and data rate for the OWC link and interrupt the communication for the 60 GHz radio link. Due to the rate-adaption, the LED-based OWC link will have a reduced data rate but the link is not interrupted, same as in in challenging weather conditions.

**Conclusions**

We will showcase an optical wireless link with Gbit/s data rates over typical distances for fixed wireless access. The design is robust against vibration and the infrastructure module can be deployed, e.g., at a lamp post with no need for tracking. In contrast to a 60 GHz link, which needs an outdoor antenna, the optical link enables an all-indoor customer premise equipment (CPE), which can be deployed by the customer without involving the house owner.

**Acknowledgement**

This work was supported by the EU in the H2020 project ELIoT under the grant agreement number 825651.